\documentclass[%
reprint,
superscriptaddress,
altaffilletter,
showpacs,
 amsmath,amssymb,
 aps,
]{revtex4-1}

\usepackage[percent]{overpic}

\usepackage{graphicx}
\usepackage{dcolumn}
\usepackage{subfigure}
\usepackage{bm}
\usepackage{multirow}

\begin{document}


\title{\textbf{Double-beta decay of $^{130}$Te to the first $0^{+}$ excited state of $^{130}$Xe with CUORICINO}}

\author{E.~Andreotti}
\altaffiliation[Presently at: ]{Joint Research Center, Institute for Reference Materials and Measurement, 2440 Geel - Belgium}
\affiliation{Dipartimento di Fisica e Matematica, Universit\`{a} dell'Insubria, Como I-22100 - Italy}
\affiliation{ INFN - Sezione di Milano Bicocca, Milano I-20126 - Italy}

\author{C.~Arnaboldi}
\affiliation{Dipartimento di Fisica, Universit\`{a} di Milano-Bicocca, Milano I-20126 - Italy}

\author{F.~T.~Avignone~III}
\affiliation{Department of Physics and Astronomy, University of South Carolina, Columbia, SC 29208 - USA}

\author{M.~Balata}
\affiliation{INFN - Laboratori Nazionali del Gran Sasso, Assergi (L'Aquila) I-67010 - Italy}

\author{I.~Bandac}
\affiliation{Department of Physics and Astronomy, University of South Carolina, Columbia, SC 29208 - USA}

\author{M.~Barucci}
\affiliation{Dipartimento di Fisica, Universit\`{a} di Firenze, Firenze I-50125 - Italy}
\affiliation{INFN - Sezione di Firenze, Firenze I-50125 - Italy}

\author{J.~W.~Beeman}
\affiliation{Materials Science Division, Lawrence Berkeley National Laboratory, Berkeley, CA 94720 - USA}

\author{F.~Bellini}
\affiliation{Dipartimento di Fisica, Sapienza Universit\`{a} di Roma, Roma  I-00185 - Italy}
\affiliation{INFN - Sezione di Roma, Roma I-00185 - Italy}

\author{C.~Brofferio}
\affiliation{ INFN - Sezione di Milano Bicocca, Milano I-20126 - Italy}
\affiliation{Dipartimento di Fisica, Universit\`{a} di Milano-Bicocca, Milano I-20126 - Italy}

\author{A.~Bryant}
\affiliation{Nuclear Science Division, Lawrence Berkeley National Laboratory, Berkeley, CA 94720 - USA}
\affiliation{Department of Physics, University of California, Berkeley, CA 94720 - USA}

\author{C.~Bucci}
\affiliation{INFN - Laboratori Nazionali del Gran Sasso, Assergi (L'Aquila) I-67010 - Italy}

\author{L.~Canonica}
\affiliation{Dipartimento di Fisica, Universit\`{a} di Genova, Genova I-16146 - Italy}
\affiliation{INFN - Sezione di Genova, Genova I-16146 - Italy}

\author{S.~Capelli}
\affiliation{ INFN - Sezione di Milano Bicocca, Milano I-20126 - Italy}
\affiliation{Dipartimento di Fisica, Universit\`{a} di Milano-Bicocca, Milano I-20126 - Italy}

\author{L.~Carbone}
\affiliation{ INFN - Sezione di Milano Bicocca, Milano I-20126 - Italy}

\author{M.~Carrettoni}
\affiliation{ INFN - Sezione di Milano Bicocca, Milano I-20126 - Italy}
\affiliation{Dipartimento di Fisica, Universit\`{a} di Milano-Bicocca, Milano I-20126 - Italy}

\author{M.~Clemenza}
\affiliation{ INFN - Sezione di Milano Bicocca, Milano I-20126 - Italy}
\affiliation{Dipartimento di Fisica, Universit\`{a} di Milano-Bicocca, Milano I-20126 - Italy}

\author{O.~Cremonesi}
\affiliation{ INFN - Sezione di Milano Bicocca, Milano I-20126 - Italy}

\author{R.~J.~Creswick}
\affiliation{Department of Physics and Astronomy, University of South Carolina, Columbia, SC 29208 - USA}

\author{S.~Di~Domizio}
\affiliation{Dipartimento di Fisica, Universit\`{a} di Genova, Genova I-16146 - Italy}
\affiliation{INFN - Sezione di Genova, Genova I-16146 - Italy}

\author{M.~J.~Dolinski}
\affiliation{Department of Physics, University of California, Berkeley, CA 94720 - USA}
\affiliation{Lawrence Livermore National Laboratory, Livermore, CA 94550 - USA}

\author{L.~Ejzak}
\affiliation{Department of Physics, University of Wisconsin, Madison, WI 53706 - USA}

\author{R.~Faccini}
\affiliation{Dipartimento di Fisica, Sapienza Universit\`{a} di Roma, Roma  I-00185 - Italy}
\affiliation{INFN - Sezione di Roma, Roma I-00185 - Italy}

\author{H.~A.~Farach}
\affiliation{Department of Physics and Astronomy, University of South Carolina, Columbia, SC 29208 - USA}

\author{E.~Ferri}
\affiliation{ INFN - Sezione di Milano Bicocca, Milano I-20126 - Italy}
\affiliation{Dipartimento di Fisica, Universit\`{a} di Milano-Bicocca, Milano I-20126 - Italy}

\author{E.~Fiorini}
\thanks{Corresponding author}
\email[e-mail: ]{Ettore.Fiorini@mib.infn.it}
\affiliation{ INFN - Sezione di Milano Bicocca, Milano I-20126 - Italy}
\affiliation{Dipartimento di Fisica, Universit\`{a} di Milano-Bicocca, Milano I-20126 - Italy}

\author{L.~Foggetta}
\altaffiliation[Presently at: ]{Laboratoire de l'Acc\'{e}l\'{e}rateur Lin\'{e}aire, Centre Scientifique d'Orsay, 91898 Orsay - France}
\affiliation{Dipartimento di Fisica e Matematica, Universit\`{a} dell'Insubria, Como I-22100 - Italy}
\affiliation{ INFN - Sezione di Milano Bicocca, Milano I-20126 - Italy}

\author{A.~Giachero}
\affiliation{ INFN - Sezione di Milano Bicocca, Milano I-20126 - Italy}

\author{L.~Gironi}
\affiliation{ INFN - Sezione di Milano Bicocca, Milano I-20126 - Italy}
\affiliation{Dipartimento di Fisica, Universit\`{a} di Milano-Bicocca, Milano I-20126 - Italy}

\author{A.~Giuliani}
\altaffiliation[Presently at: ]{Centre de Spectrom\'{e}trie Nucl\'{e}aire et de Spectrom\'{e}trie de Masse, 91405 Orsay Campus - France}
\affiliation{Dipartimento di Fisica e Matematica, Universit\`{a} dell'Insubria, Como I-22100 - Italy}
\affiliation{ INFN - Sezione di Milano Bicocca, Milano I-20126 - Italy}

\author{P.~Gorla}
\altaffiliation[Presently at: ]{INFN - Sezione di Roma Tor Vergata, Roma I-00133 - Italy}
\affiliation{INFN - Laboratori Nazionali del Gran Sasso, Assergi (L'Aquila) I-67010 - Italy}

\author{E.~Guardincerri}
\affiliation{INFN - Laboratori Nazionali del Gran Sasso, Assergi (L'Aquila) I-67010 - Italy}
\affiliation{Nuclear Science Division, Lawrence Berkeley National Laboratory, Berkeley, CA 94720 - USA}
\affiliation{INFN - Sezione di Genova, Genova I-16146 - Italy}

\author{T.~D.~Gutierrez}
\affiliation{Physics Department, California Polytechnic State University, San Luis Obispo, CA 93407 - USA}

\author{E.~E.~Haller}
\affiliation{Materials Science Division, Lawrence Berkeley National Laboratory, Berkeley, CA 94720 - USA}
\affiliation{Department of Materials Science and Engineering, University of California, Berkeley, CA 94720 - USA}

\author{K.~Kazkaz}
\affiliation{Lawrence Livermore National Laboratory, Livermore, CA 94550 - USA}

\author{L.~Kogler}
\affiliation{Nuclear Science Division, Lawrence Berkeley National Laboratory, Berkeley, CA 94720 - USA}
\affiliation{Department of Physics, University of California, Berkeley, CA 94720 - USA}

\author{S.~Kraft}
\affiliation{ INFN - Sezione di Milano Bicocca, Milano I-20126 - Italy}
\affiliation{Dipartimento di Fisica, Universit\`{a} di Milano-Bicocca, Milano I-20126 - Italy}

\author{C.~Maiano}
\affiliation{ INFN - Sezione di Milano Bicocca, Milano I-20126 - Italy}
\affiliation{Dipartimento di Fisica, Universit\`{a} di Milano-Bicocca, Milano I-20126 - Italy}

\author{C.~Martinez}
\altaffiliation[Presently at: ]{Queen's University, Kingston, ON K7L 3N6 - Canada}
\affiliation{Department of Physics and Astronomy, University of South Carolina, Columbia, SC 29208 - USA}

\author{M.~Martinez}
\altaffiliation[Presently at: ]{Institut d'Astrophysique Spatial, 91045 Orsay - France}
\affiliation{ INFN - Sezione di Milano Bicocca, Milano I-20126 - Italy}
\affiliation{Laboratorio de Fisica Nuclear y Astroparticulas, Universidad de Zaragoza, Zaragoza 50009 - Spain}

\author{R.~H.~Maruyama}
\affiliation{Department of Physics, University of Wisconsin, Madison, WI 53706 - USA}

\author{S.~Newman}
\affiliation{Department of Physics and Astronomy, University of South Carolina, Columbia, SC 29208 - USA}
\affiliation{INFN - Laboratori Nazionali del Gran Sasso, Assergi (L'Aquila) I-67010 - Italy}

\author{S.~Nisi}
\affiliation{INFN - Laboratori Nazionali del Gran Sasso, Assergi (L'Aquila) I-67010 - Italy}

\author{C.~Nones}
\altaffiliation[Presently at: ]{CEA / Saclay, 91191 Gif-sur-Yvette - France}
\affiliation{Dipartimento di Fisica e Matematica, Universit\`{a} dell'Insubria, Como I-22100 - Italy}
\affiliation{ INFN - Sezione di Milano Bicocca, Milano I-20126 - Italy}

\author{E.~B.~Norman}
\affiliation{Lawrence Livermore National Laboratory, Livermore, CA 94550 - USA}
\affiliation{Department of Nuclear Engineering, University of California, Berkeley, CA 94720 - USA}

\author{A.~Nucciotti}
\affiliation{ INFN - Sezione di Milano Bicocca, Milano I-20126 - Italy}
\affiliation{Dipartimento di Fisica, Universit\`{a} di Milano-Bicocca, Milano I-20126 - Italy}

\author{F.~Orio}
\affiliation{Dipartimento di Fisica, Sapienza Universit\`{a} di Roma, Roma  I-00185 - Italy}
\affiliation{INFN - Sezione di Roma, Roma I-00185 - Italy}

\author{M.~Pallavicini}
\affiliation{Dipartimento di Fisica, Universit\`{a} di Genova, Genova I-16146 - Italy}
\affiliation{INFN - Sezione di Genova, Genova I-16146 - Italy}

\author{V.~Palmieri}
\affiliation{INFN - Laboratori Nazionali di Legnaro, Legnaro (Padova) I-35020 - Italy} 

\author{L.~Pattavina}
\affiliation{ INFN - Sezione di Milano Bicocca, Milano I-20126 - Italy}
\affiliation{Dipartimento di Fisica, Universit\`{a} di Milano-Bicocca, Milano I-20126 - Italy}

\author{M.~Pavan}
\affiliation{ INFN - Sezione di Milano Bicocca, Milano I-20126 - Italy}
\affiliation{Dipartimento di Fisica, Universit\`{a} di Milano-Bicocca, Milano I-20126 - Italy}

\author{M.~Pedretti}
\affiliation{Lawrence Livermore National Laboratory, Livermore, CA 94550 - USA}

\author{G.~Pessina}
\affiliation{ INFN - Sezione di Milano Bicocca, Milano I-20126 - Italy}

\author{S.~Pirro}
\affiliation{ INFN - Sezione di Milano Bicocca, Milano I-20126 - Italy}

\author{E.~Previtali}
\affiliation{ INFN - Sezione di Milano Bicocca, Milano I-20126 - Italy}

\author{L.~Risegari}
\affiliation{Dipartimento di Fisica, Universit\`{a} di Firenze, Firenze I-50125 - Italy}
\affiliation{INFN - Sezione di Firenze, Firenze I-50125 - Italy}

\author{C.~Rosenfeld}
\affiliation{Department of Physics and Astronomy, University of South Carolina, Columbia, SC 29208 - USA}

\author{C.~Rusconi}
\affiliation{Dipartimento di Fisica e Matematica, Universit\`{a} dell'Insubria, Como I-22100 - Italy}
\affiliation{ INFN - Sezione di Milano Bicocca, Milano I-20126 - Italy}

\author{C.~Salvioni}
\affiliation{Dipartimento di Fisica e Matematica, Universit\`{a} dell'Insubria, Como I-22100 - Italy}
\affiliation{ INFN - Sezione di Milano Bicocca, Milano I-20126 - Italy}

\author{S.~Sangiorgio}
\altaffiliation[Presently at: ]{Lawrence Livermore National Laboratory, Livermore, CA 94550 - USA}
\affiliation{Department of Physics, University of Wisconsin, Madison, WI 53706 - USA}

\author{D.~Schaeffer}
\affiliation{ INFN - Sezione di Milano Bicocca, Milano I-20126 - Italy}
\affiliation{Dipartimento di Fisica, Universit\`{a} di Milano-Bicocca, Milano I-20126 - Italy}

\author{N.~D.~Scielzo}
\affiliation{Lawrence Livermore National Laboratory, Livermore, CA 94550 - USA}

\author{M.~Sisti}
\affiliation{ INFN - Sezione di Milano Bicocca, Milano I-20126 - Italy}
\affiliation{Dipartimento di Fisica, Universit\`{a} di Milano-Bicocca, Milano I-20126 - Italy}

\author{A.~R.~Smith}
\affiliation{EH\&S Division, Lawrence Berkeley National Laboratory, Berkeley, CA 94720 - USA}

\author{C.~Tomei}
\affiliation{INFN - Sezione di Roma, Roma I-00185 - Italy}

\author{G.~Ventura}
\affiliation{Dipartimento di Fisica, Universit\`{a} di Firenze, Firenze I-50125 - Italy}
\affiliation{INFN - Sezione di Firenze, Firenze I-50125 - Italy}

\author{M.~Vignati}
\affiliation{Dipartimento di Fisica, Sapienza Universit\`{a} di Roma, Roma  I-00185 - Italy}
\affiliation{INFN - Sezione di Roma, Roma I-00185 - Italy}

\date{\today}

\begin{abstract}
The CUORICINO experiment was an array of 62 TeO$_{2}$ single-crystal bolometers with a total $^{130}$Te mass of $11.3\,$kg.
The experiment finished in 2008 after more than 3 years of active operating time. 
Searches for both $0\nu$ and $2\nu$ double-beta decay to the first excited $0^{+}$ state in $^{130}$Xe were performed by studying different coincidence scenarios.
The analysis was based on data representing a total exposure of N($^{130}$Te)$\cdot$t=$9.5\times10^{25}\,$y.
No evidence for a signal was found.
The resulting lower limits on the half lives are $T^{2\nu}_{\frac{1}{2}}\left(^{130}\mbox{Te}\rightarrow^{130}\mbox{Xe}^{*}\right)>1.3\times10^{23}\,$y (90\% C.L.), and $T^{0\nu}_{\frac{1}{2}}\left(^{130}\mbox{Te}\rightarrow^{130}\mbox{Xe}^{*}\right)>9.4\times10^{23}\,$y (90\% C.L.).
\end{abstract}

\maketitle


\section{Introduction}\label{sec:intro}

Two-neutrino double-beta ($2\nu\beta\beta$) decay and neutrinoless double-beta ($0\nu\beta\beta$) decay have been known for over 70 years now \cite{goeppert1935, racah1937} (a recent review can be found in \cite{avignone2008}).
While experimental evidence for $2\nu\beta\beta$-decay has been found there is still no observation for the $0\nu\beta\beta$-decay, however several limits for the half-life have been set in the past with values greater than $10^{21}\,$y.
In both of these processes the lifetime is proportional to the square of the Nuclear Matrix Elements (NME).
Two neutrino double beta decay has been detected in ten nuclei on the ground state of the daughter nucleus and in two nuclei on the excited state of it, and the corresponding extracted values for the NME are in reasonable agreement with the theoretical expectation.
In the case of $0\nu\beta\beta$-decay their value is very important since it plays the same role in the prediction of the decay time as $m_{\beta\beta}$, the effective neutrino mass \cite{avignone2008, rodin2006, rodin2007,medex}.

\begin{figure*}[tb]
\begin{center}
    \includegraphics[width=.95\textwidth]{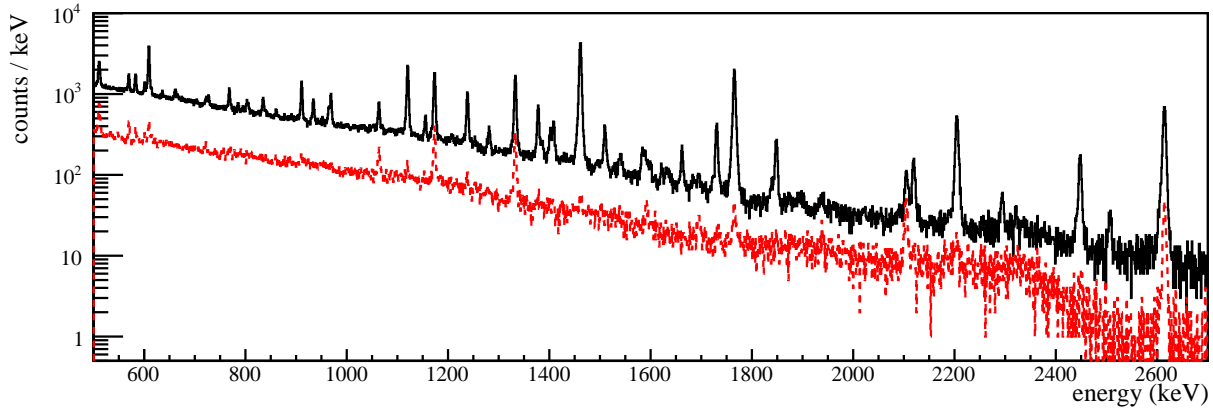}
\caption{Single-hit (black line) and double-hit (red dashed line) energy spectra collected by CUORICINO in the range (500$\div$2700)$\,$keV.}
\label{fig:cuoricino_spectrum}
\end{center}
\end{figure*}

The CUORICINO experiment was an array of 62 TeO$_{2}$ bolometers operated at a temperature of about $10\,$mK.
A bolometer~\cite{bolometers:enss, bolometers:booth} detects an energy release as a temperature rise in the absorber crystal.
Thermal pulses are converted into electric signals by means of neutron transmutation doped (NTD) thermistors~\cite{ntd}, which are coupled to each absorber.
CUORICINO was organized in 13 planes.
All of these planes were composed of four crystals with dimensions of $5\times5\times5\,$cm$^{3}$ and a mass of $790\,$g each, except for the $11^{th}$ and $12^{th}$ (from top to bottom).
Each of these two particular planes had 9 crystals with dimensions of $3\times3\times6\,$cm$^{3}$ and a mass of $330\,$g.
Two of these smaller crystals were enriched to 82.3$\%$ of $^{128}$Te and two others to 75$\%$ of $^{130}$Te.
All the other crystals had the natural isotopic abundance of $^{130}$Te (33.8\%).
A monthly calibration was performed using a $^{232}$Th source.
The energy spectrum of the events collected by CUORICINO can be seen in Figure~\ref{fig:cuoricino_spectrum}.
A more detailed description of the experiment can be found in~\cite{arnaboldi2008}.

\begin{figure}[pt]
\begin{center}
\includegraphics[width=.48\textwidth]{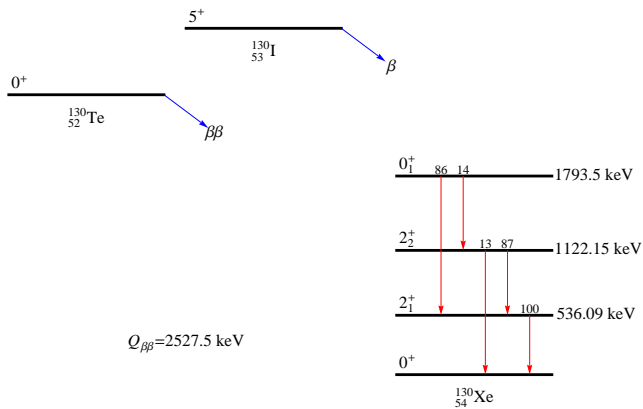}
\caption{Decay scheme for $^{130}$Te, showing the energy levels (keV) and the branching ratios for the $\gamma$-rays~\cite{SINGH200133}.}
\label{fig:dScheme}
\end{center}
\end{figure}

CUORICINO's geometry provides a unique opportunity to search for $0\nu\beta\beta$ and $2\nu\beta\beta$ decay to the first $0^{+}$ excited state in $^{130}$Xe in an essentially background-free environment.
This is due to the fact that these processes can be studied using a coincidence-based analysis by searching for two $\gamma$ lines of well defined energy.
As can be seen in Figure~\ref{fig:dScheme}, the decay to the first $0^{+}$ excited state in $^{130}$Xe differs from the one to the ground state in that it produces a gamma cascade.
Given the Q-value of the decay, $Q_{\beta\beta}$=2527.5$\,$keV~\cite{Redshaw:2009zz, Scielzo:2009nh, Rahaman2011}, the two electrons are left with a total energy of 734.0$\,$keV.
The most probable de-excitation pattern, with a 86\% branching ratio, 
proceeds through the emission of a $1257.41\,$keV and a $536.09\,$keV gamma.
Though $2\nu\beta\beta$ and $0\nu\beta\beta$ decay both result in the emission of two electrons, the spectra of the sum energy of the two electrons differ drastically.
In the first case, the two resulting betas have a continuous spectrum in the range (0$\div$734.0)$\,$keV, while in the second case, the result is just a monochromatic beta peak centered at 734.0$\,$keV.
Theoretical evaluations and experimental limits for these two processes can be found in Table~\ref{tab:prevpub}.
It is important to note that the theoretical calculation for the half life of 2$\nu\beta\beta$-decay to the first excited state 0$^+$ reported in Table~\ref{tab:prevpub} is not the one originally indicated in reference~\cite{suhonen1997} since it was based on a wrong evaluation of the phase-space.
The reported  value is the one re-elaborated by A.S. Barabash~\cite{barabash2001} on the basis of  the correct phase space factor.
\begin{table}[tb]
\begin{center}
\begin{tabular}{cccc}
&\\
\hline\hline
Decay & Transition & Theoretical (y) & Experimental (y)\\
\hline
\multirow{2}{*}{0$\nu$} & $0^{+}\rightarrow0_{1}^{+}$ & 7.5$\times$10$^{25}$~\cite{suhonen2000, suhonen2003} & $>3.1\times$10$^{22}$~\cite{Arnaboldi2003}\\
 & $0^{+}\rightarrow0^{+}$ & (1.6$\div$15)$\times$10$^{23}$~\cite{Tretyak:2002dx}& $>2.8\times$10$^{24}$~\cite{Andreotti:2010vj} \\
\hline
\multirow{2}{*}{2$\nu$} & $0^{+}\rightarrow0_{1}^{+}$ & (5.1$\div$14)$\times$10$^{22}$~\cite{barabash2001, suhonen1997}\footnotemark[1] & $>2.3\times$10$^{21}$~\cite{barabash2001}\\
 & $0^{+}\rightarrow0^{+}$ & (1.7$\div$70)$\times$10$^{19}$~\cite{Tretyak:2002dx} & $7.0\times$10$^{20}$~\cite{Arnold:2011gq} \\
\hline\hline
\end{tabular}
\footnotetext[1]{Corrected values for \cite{suhonen1997} (discussion in text)}
\caption{Theoretical evaluations (for $m_{\beta\beta}$=1$\,$eV) and experimental best limits (90\% CL) for the half-life of $^{130}$Te $0\nu\beta\beta$ and $2\nu\beta\beta$ decay.}
\label{tab:prevpub}
\end{center}
\end{table}

\section{Search Strategy and Event Selection}\label{sec:expro}

\begin{figure}[tb]
 \begin{center}
     	\subfigure[Scenario 1]{
	\includegraphics[width=.2\textwidth]{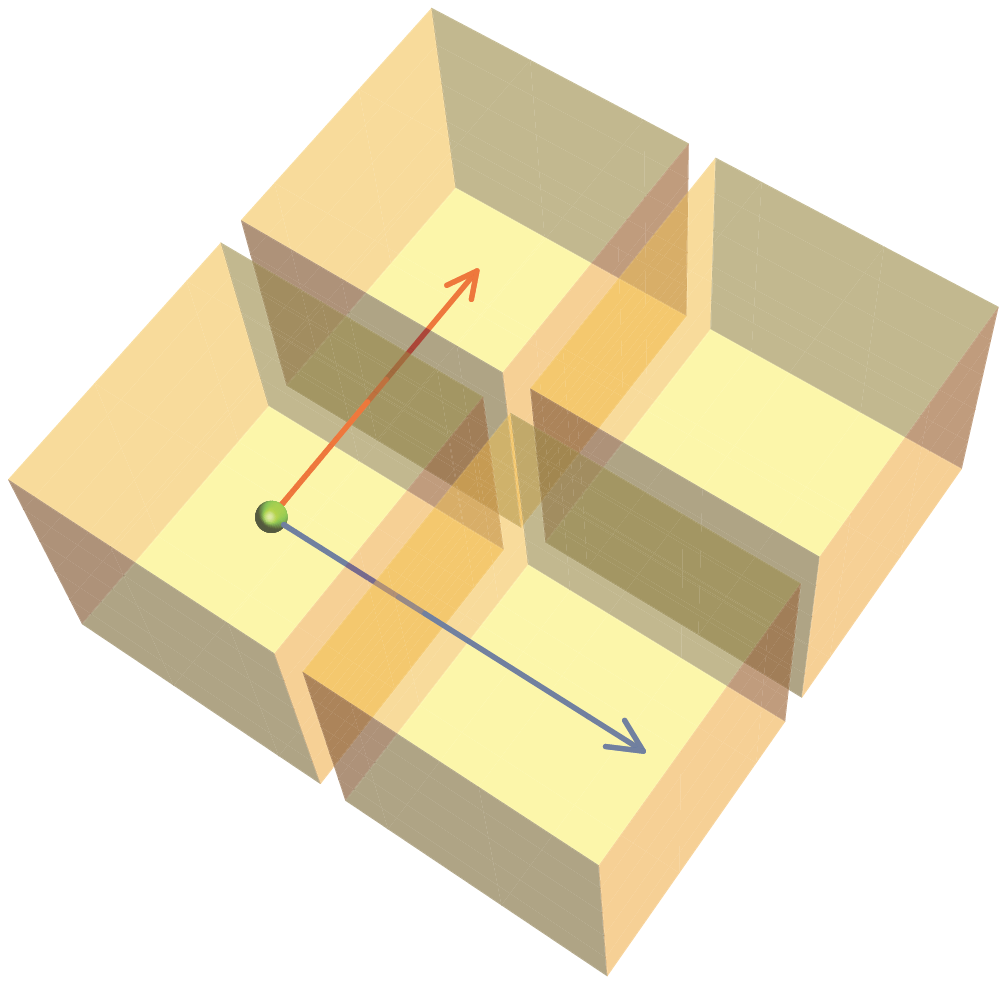}
	\label{pScenarios1}}
  	\subfigure[Scenario 2]{
     	\includegraphics[width=.2\textwidth]{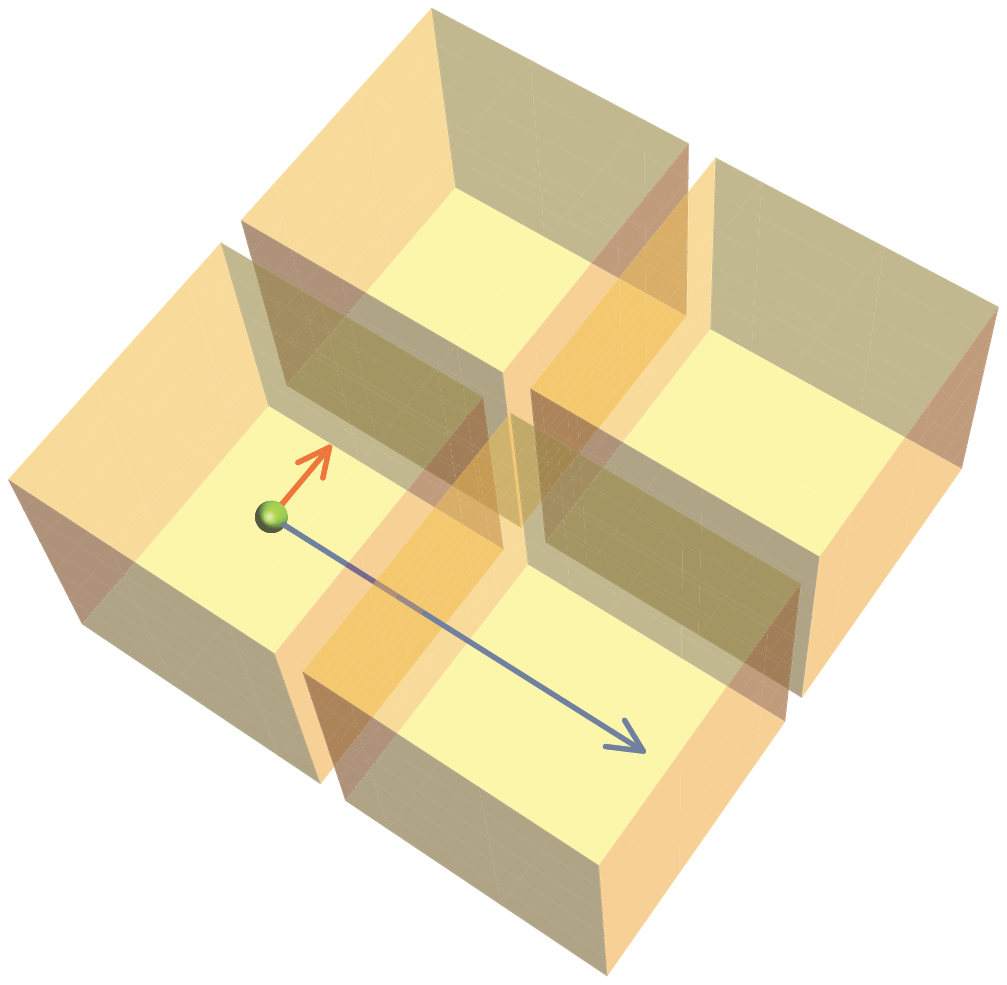}
    	\label{pScenarios2}}\\
   	\subfigure[Scenario 3]{
    	\includegraphics[width=.2\textwidth]{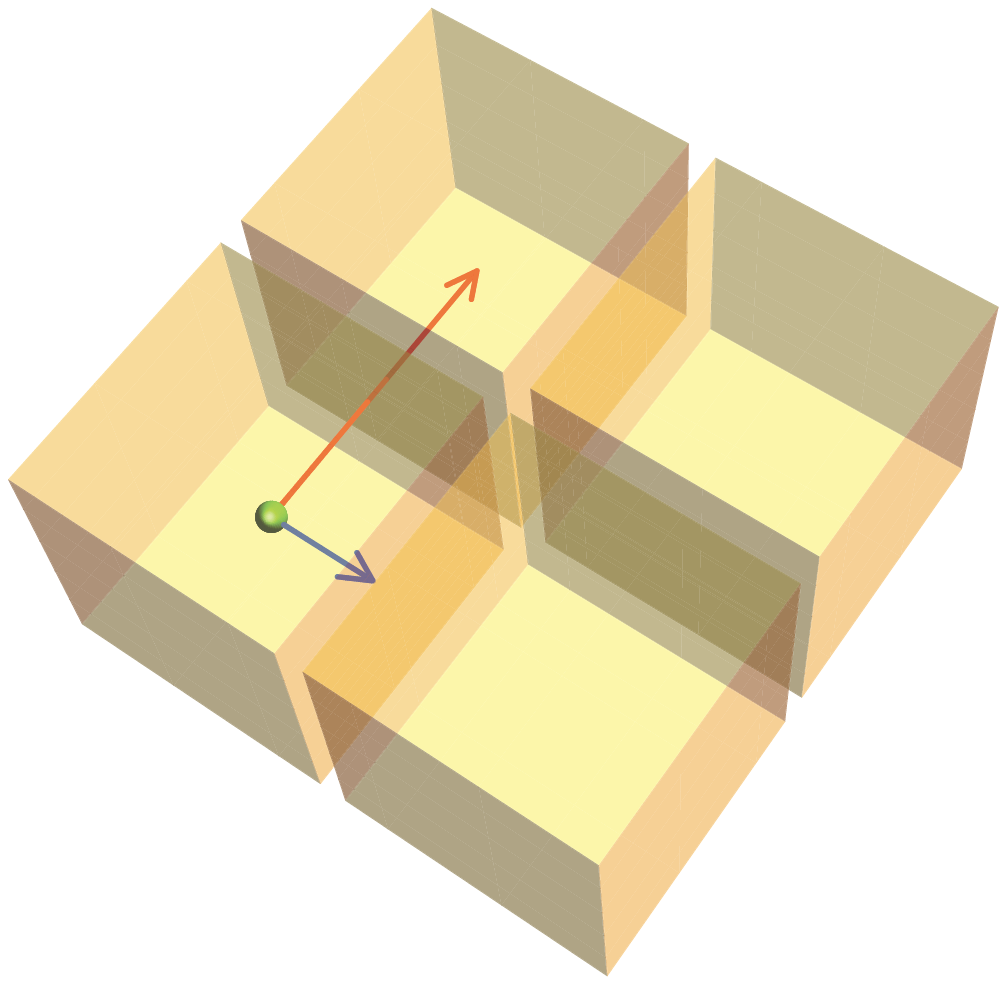}
     	\label{pScenarios3}}
 \end{center}
 \caption{Possible capture scenarios. The blue lines represent the 1257.41$\,$keV $\gamma$, while the red lines represent the 536.09$\,$keV one. For each scenario, the available energy for the emitted $\beta$s is 734.0$\,$keV.}
 \label{fig:pScenarios}
 \end{figure}

In this analysis, we consider only configurations in which the electrons are contained in the crystal where the decay takes place, and each de-excitation photon is completely absorbed in one crystal.
With these requirements, three different scenarios are possible (see Figure~\ref{fig:pScenarios}).
Scenario 1 takes place when both gammas escape from the original crystal.
In scenario 2, the low-energy gamma (536.09$\,$keV) is trapped in the original crystal with the betas, while the high-energy gamma (1257.41$\,$keV) escapes. Scenario 3 is the opposite of scenario 2: the high-energy gamma (1257.41$\,$keV) is trapped in the original crystal with the betas, while the low-energy one (536.09$\,$keV) escapes.
The signatures and the corresponding efficiencies are reported in Table~\ref{tab:cuts}.
A further explanation of the calculation of the efficiencies can be found in Section~\ref{sec:anre}.

\begin{table*}[tb]
\begin{center}
\begin{tabular}{cccccc}
\hline\hline
Decay mode & Scenario & Signature (energies in keV) & \multicolumn{3}{c}{Efficiency}\\
 &	&	& MC & Instrumental & Total\\
\hline
\multirow{3}{*}{$0\nu$}  & 1 & 734~($\beta$) + 536~($\gamma$)  + 1257~($\gamma$) & (0.60$\pm$0.02)\% & (86$\pm$2)\% & (0.44 $\pm$ 0.02)\%\\
& 2 & 1257~($\gamma$) + 1270~($\beta+\gamma$)       & (2.29$\pm$0.04)\% & (90$\pm$1)\% & (1.77 $\pm$ 0.04)\%\\
 & 3 & 536~($\gamma$)  + 1991~($\beta+\gamma$)       & (1.41$\pm$0.03)\% & (90$\pm$1)\% & (1.09 $\pm$ 0.03)\%\\
\hline
\multirow{3}{*}{$2\nu$}  & 1 & (0 $\div$ 734)~($\beta$)  + 536~($\gamma$) + 1257~($\gamma$) & (0.53$\pm$0.02)\% & (86$\pm$2)\% & (0.39 $\pm$ 0.02)\%\\
 & 2 & (536 $\div$ 1270)~($\beta+\gamma$)  + 1257~($\gamma$)    & (3.04$\pm$0.04)\% & (90$\pm$1)\% & (2.35 $\pm$ 0.04)\%\\
 & 3 & (1257 $\div$ 1991)~($\beta+\gamma$)  + 536~($\gamma$)   & (1.28$\pm$0.03)\% & (90$\pm$1)\% & (0.99 $\pm$ 0.03)\%\\
\hline\hline
\end{tabular}
\end{center}
\caption{Signatures and efficiencies for the three scenarios for 0$\nu$ and 2$\nu$ decay. We denote with the + sign the coincidence of energies released in different crystals. Efficiencies labeled as MC were computed based on Monte Carlo simulations. Instrumental efficiencies were computed based on CUORICINO data. Total efficiencies are given by the product of MC and instrumental efficiencies, times a factor of 0.86 to account for the branching ratio of the considered decay scheme (see Figure~\ref{fig:dScheme}).}
\label{tab:cuts}
\end{table*}

The first-level analysis of the CUORICINO data is common to all physics processes to be studied and is described in detail in~\cite{Andreotti:2010vj}.
It starts from raw events and ends with a set of energy-calibrated hits associated with a time, a crystal, and other ancillary information, such as pulse shape parameters.
In this phase of the analysis, a channel- and time-dependent energy threshold is applied to the data, based on the performance of each bolometer.

For the processes studied in this paper, the analysis consists of defining signatures according to the three scenarios reported in Table~\ref{tab:cuts},
using them to select events from the CUORICINO data and evaluating the corresponding efficiencies from GEANT4-based Monte Carlo simulations~\cite{bucci2009}.

Event selection criteria can be grouped into three categories: global, event-based and coincidence-based.
Global and event-based cuts are not specific to this analysis, and here we only outline them briefly (refer to~\cite{Andreotti:2010vj} for details).
Defined a priori, global cuts are used to discard time windows in which one or more detectors performed poorly.
This could happen because of external noise or cryogenic instabilities, which in turn result in a bad energy resolution.
Event-based cuts allow the exclusion of non-physical pulses (electronic spikes or cryogenic-induced pulses) and physical pulses for which the energy is not estimated correctly (pile-up or excessive noise superimposed on the pulse).

Coincidence-based cuts rely on the properties of a group of events that occurred within a fixed time window.
Events can be selected based on the number of involved crystals, the spatial distance among them, the sum energy or the energy of the single hits.
In this paper, a 100~ms time window was used to define coincident events.
Physical coincidences induced by $^{130}$Xe de-excitation occur on much shorter time scales, but such a large time window must be chosen to account for the slow response of the bolometers.

The coincidence-based event selection criteria were decided based on the scenarios described at the beginning of this section.
Because the two electrons emitted in the 2$\nu$ decay have a continuous spectrum, a wide energy window must be chosen for one of the crystals.
This has the effect of introducing a much bigger background than is present in the analysis of the 0$\nu$ decay mode.
As a consequence, besides the criteria reported in Table~\ref{tab:cuts}, additional restrictions were applied to the events to be included in the 2$\nu$ analysis.
To reduce random coincidences, a cut was imposed on the distance between the crystals involved in the events, as it was seen from the simulation that there is a low chance for the investigated processes to involve crystals that are far apart from each other.
The most relevant background from physical processes is due to gamma rays that undergo a Compton interaction in one crystal and are then absorbed in another crystal.
While the sum energy of these events is fixed, the energy released in each crystal has a continuous distribution.
To reduce this background, events whose sum energy fell into a window of $\pm$8$\,$keV around the most intense gamma lines (1729.60$\,$keV, 1764.49$\,$keV, 1847.42$\,$keV, 2118.5$\,$keV, 2204.21$\,$keV and 2447.86$\,$keV from $^{214}$Bi, 2505$\,$keV from $^{60}$Co and 2615$\,$keV from $^{208}$Tl) were removed.

\section{Analysis}\label{sec:anre}

As stated in Section~\ref{sec:expro}, Monte Carlo simulations were used to calculate the efficiencies for the processes studied in this paper.
This was achieved by comparing the number of events passing the coincidence cuts to the total number of simulated events.
The relatively low efficiencies reported in Table~\ref{tab:cuts} arise from the fact that most of the gammas escape the crystals undetected and are absorbed by inert materials surrounding them.
Moreover, the signatures sought only consider the case of photons that are completely absorbed in one crystal, thus rejecting events in which at least one photon is absorbed in one crystal after undergoing a Compton interaction in a different one.
The computed values reported in the last column of Table~\ref{tab:cuts} also include inefficiencies due to event-based cuts, channel- and time-dependent energy thresholds, and discarded time windows in which one or more detectors were not performing properly (global cuts).
Inefficiencies induced by channel based cuts were evaluated on the CUORICINO data in the same way discussed in~\cite{Andreotti:2010vj}.
The effect of global cuts was taken into account by removing the simulated events lying in the time windows that were discarded from the real CUORICINO data, after said time windows were rescaled by the ratio between the total duration of the simulation and the real CUORICINO live time.
The same procedure was used to associate energy thresholds to the simulation.
Because the effect of global cuts was taken into consideration when determining the efficiencies, the exposure used in this work corresponds to the complete CUORICINO statistics without any subtractions: N($^{130}$Te)$\cdot$t = $9.5\times10^{25}\,$y.

\begin{center}
\begin{figure*}[tb]
\begin{center}
    \begin{overpic}[width=.49\textwidth]{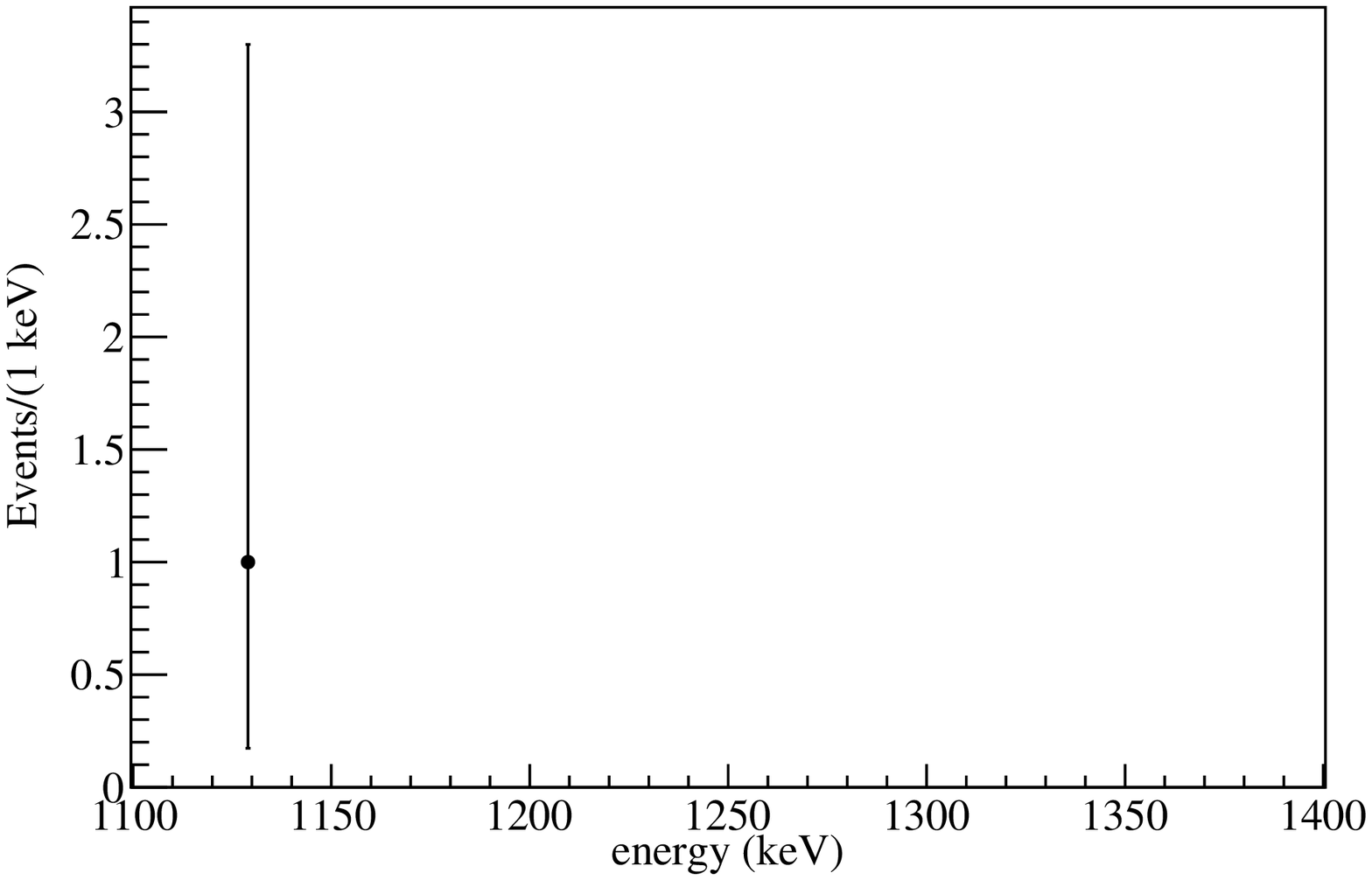}
      \put(88,56.5){A}
    \end{overpic}
    \begin{overpic}[width=.49\textwidth]{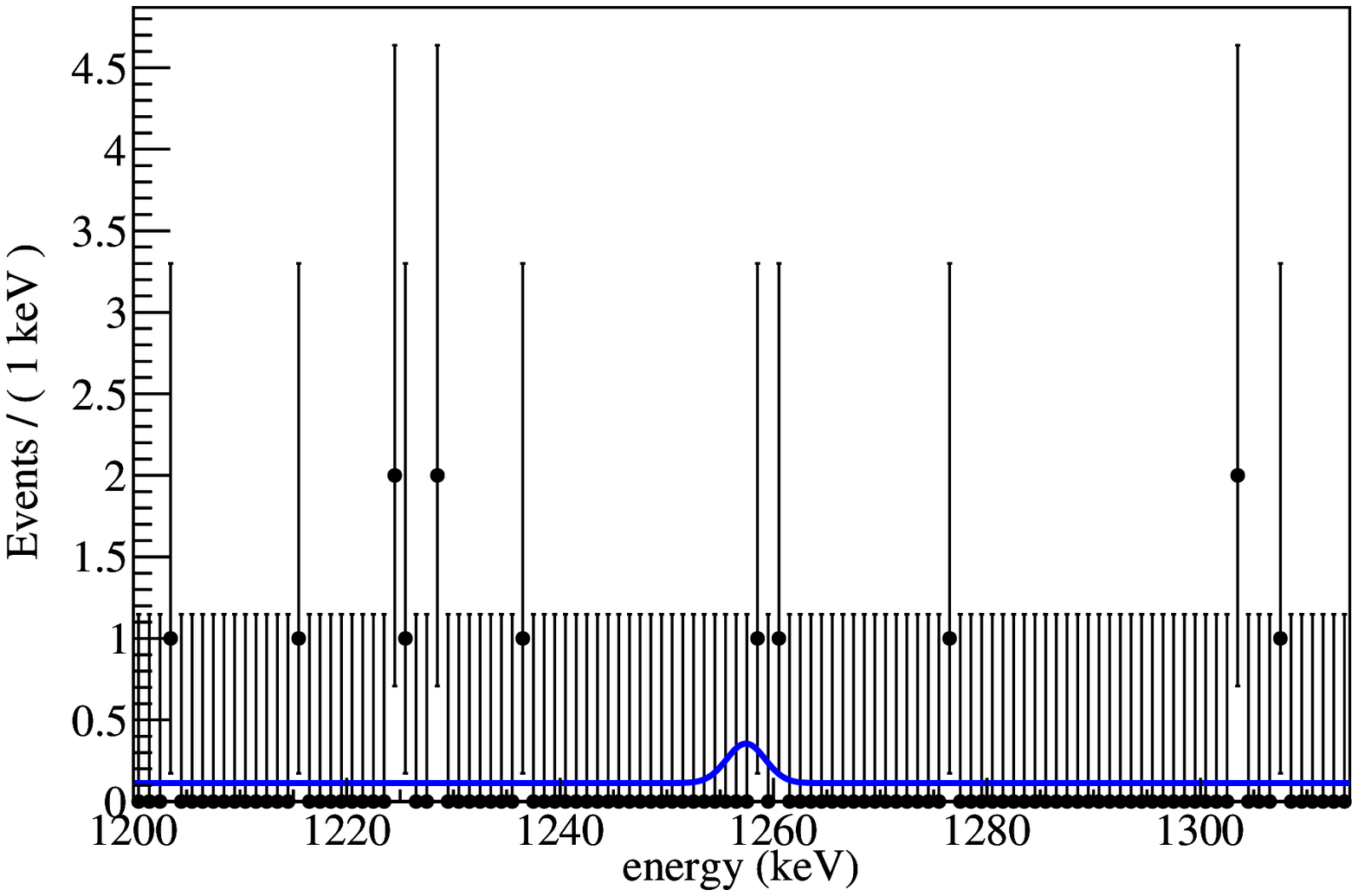}
      \put(88.5,56.5){D}
    \end{overpic}
    \begin{overpic}[width=.49\textwidth]{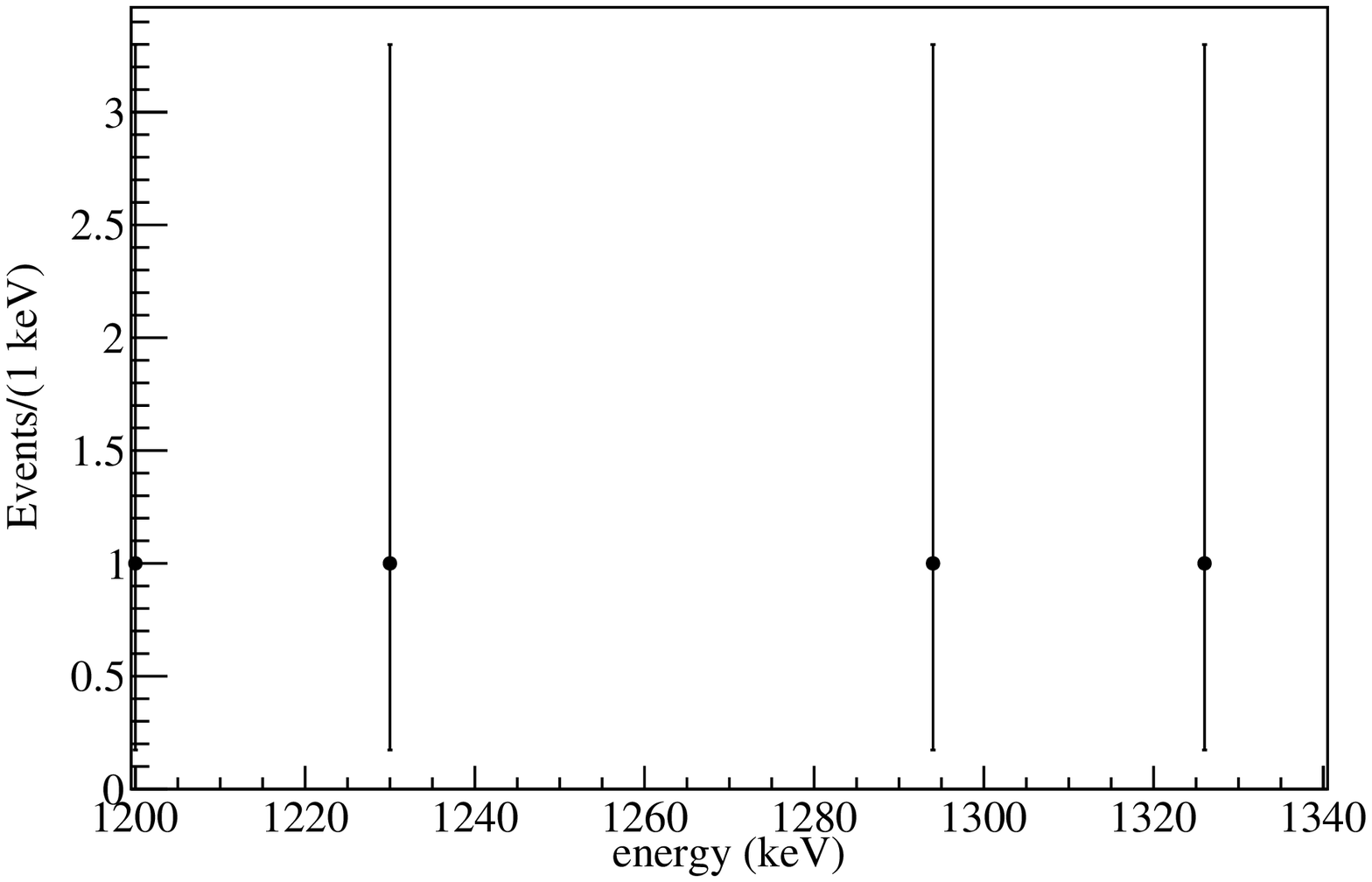}
      \put(88,56.5){B}
    \end{overpic}
    \begin{overpic}[width=.49\textwidth]{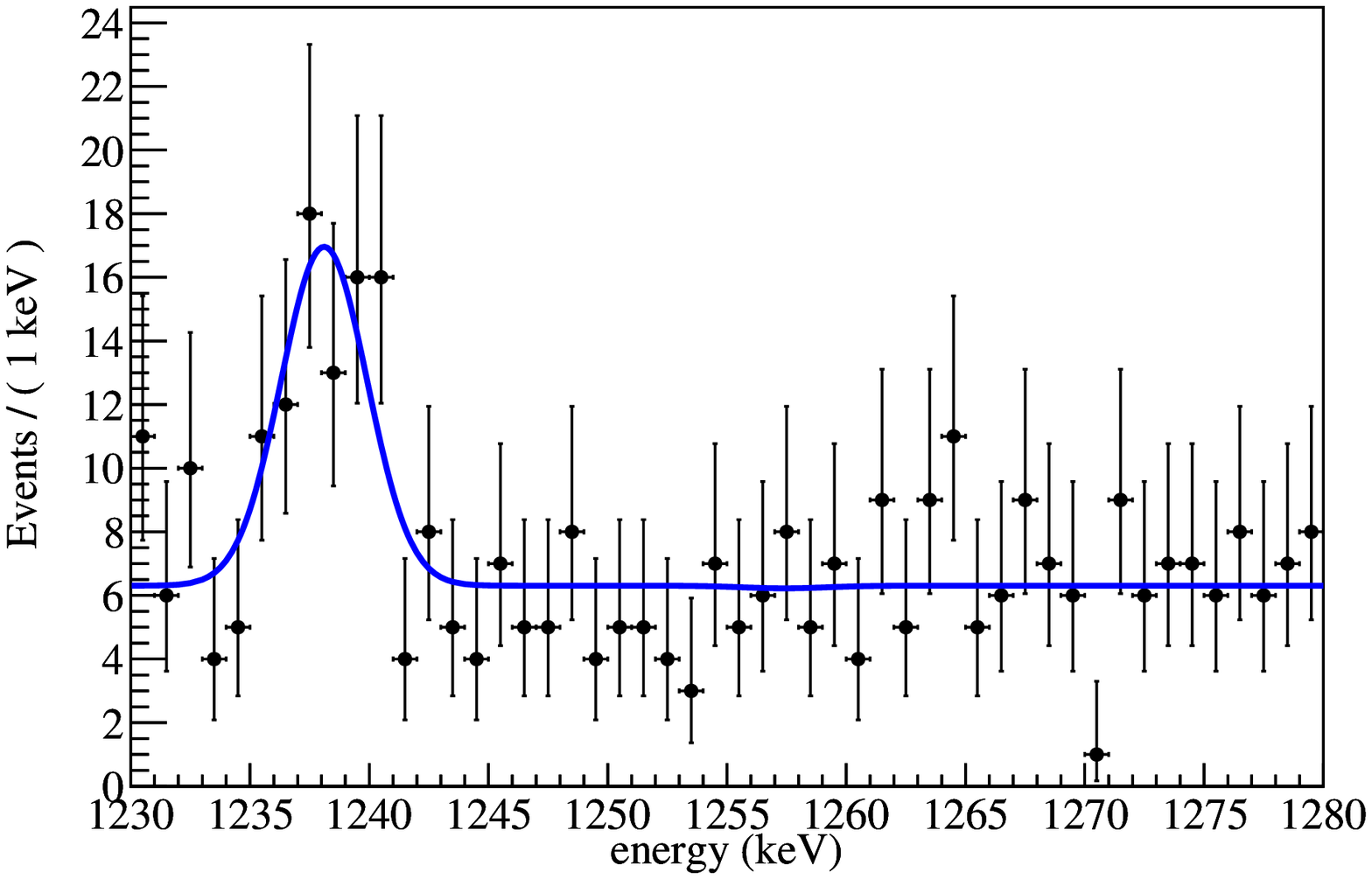}
      \put(88,56.5){E}
    \end{overpic}
    \begin{overpic}[width=.49\textwidth]{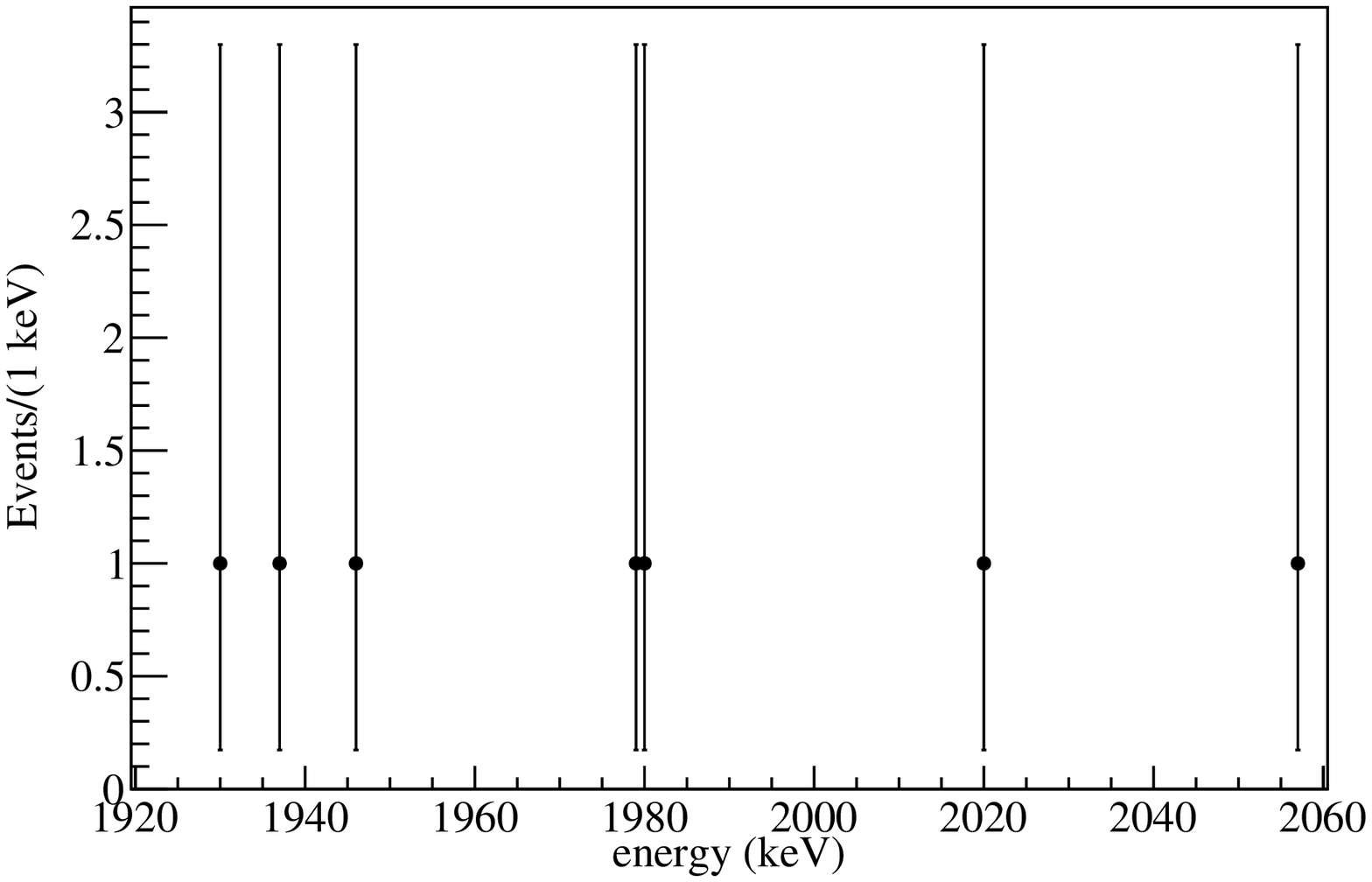}
      \put(88,56.5){C}
    \end{overpic}
    \begin{overpic}[width=.49\textwidth]{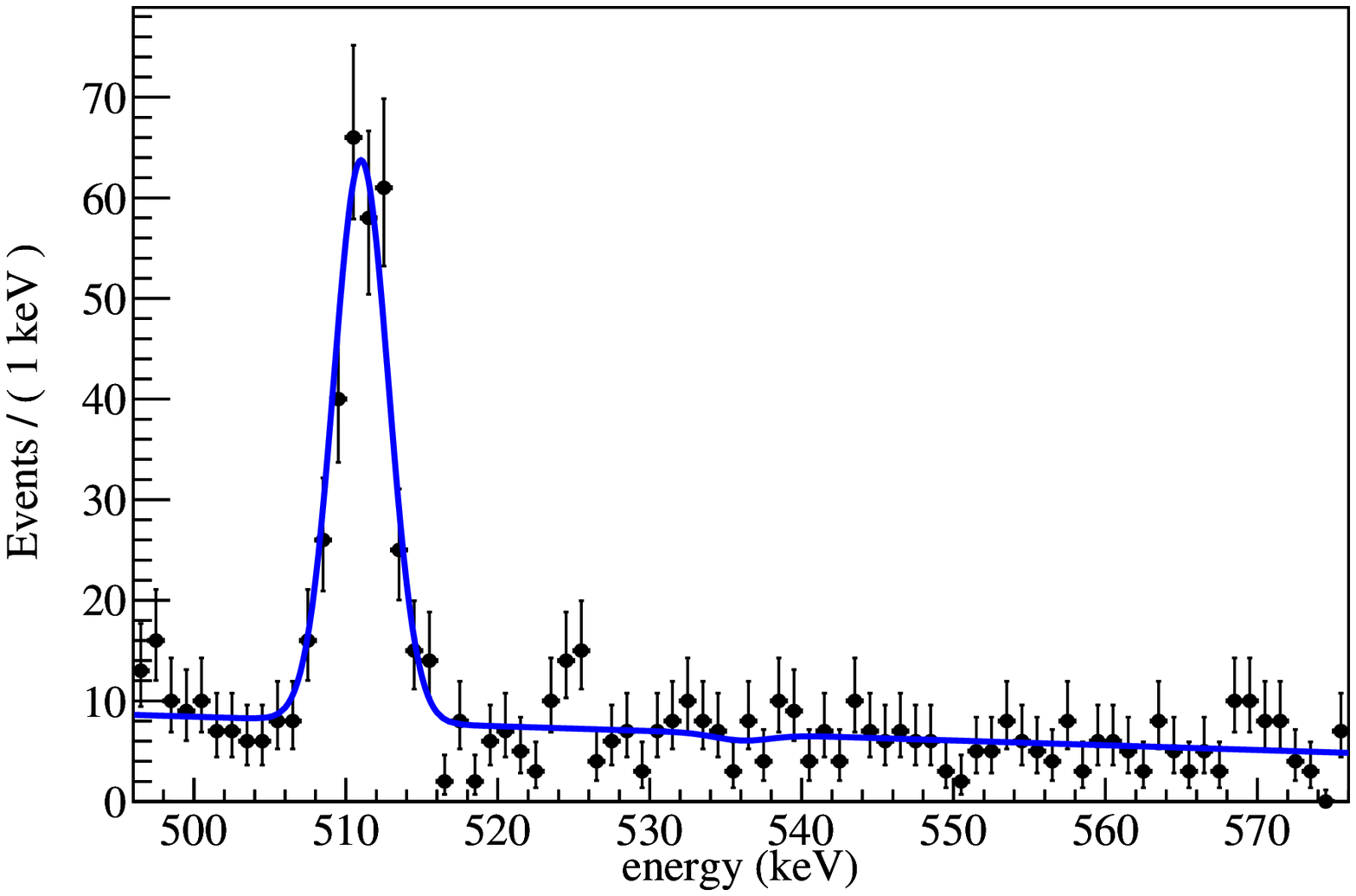}
      \put(88,56.5){F}
    \end{overpic}
\end{center}
\caption{CUORICINO energy spectra after the event selection cuts applied for the 0$\nu$ (left) and 2$\nu$ (right) analyses.
For the $0\nu$ decay, the signal was expected at 1257.41$\,$keV (plot A), 1270$\,$keV (plot B) and 1991$\,$keV (plot C) for scenarios 1, 2 and 3 respectively. For the $2\nu$ decay, the signal was expected at 1257.41$\,$keV for scenarios 1 (plot D) and 2 (plot E) and at 536.09$\,$keV for scenario 3 (plot F).}
\label{fig:spectra}
\end{figure*}
\end{center}

Figure~\ref{fig:spectra} shows the energy spectra obtained from the CUORICINO data after applying the event selection cuts described in Section~\ref{sec:expro}. For each scenario, the spectrum was built as follows.
Coincidence cuts were applied based on Table~\ref{tab:cuts}, requiring that the accepted events be in coincidence with events satisfying each component hit of the signature \emph{except for} the hit corresponding to the highest-energy $\gamma$.
The signal search could then consist of a search in the resultant spectrum for evidence of the highest-energy $\gamma$ of the signature, which is the component with the lowest background.
Moreover, the acceptance width for each cut was enlarged by $\pm$10$\,$keV with respect to the energies and energy ranges listed in Table~\ref{tab:cuts}, to account for the finite energy resolution of the detectors ($\sigma\simeq 2\,$keV; see discussion below).
The energy windows used for the spectra were chosen to be much larger than the detector resolution, but small enough that at most one radioactive background peak was included, and the continuum could be assumed to be flat or linear.

No evidence for a signal was found in any of the energy spectra.
For the zero-neutrino decay mode, the background is negligible, and no fit was performed.
In this case, a condition of zero signal and zero background was assumed.
In contrast, the background is not negligible for the 2$\nu$ decay mode, and therefore a Bayesian maximum likelihood fit was performed for the 2$\nu$ analyses.
The best-fit curves are represented by the blue lines in Figure~\ref{fig:spectra}.
Depending on the scenario, different background models were adopted for the 2$\nu$ spectra.
The continuum was fitted with a constant (scenarios 1 and 2) or linear shape (scenario 3), while the possible additional peaks (1238$\,$keV from $^{214}$Bi for scenario~2, 511$\,$keV for scenario 3) were fitted with a Gaussian shape.
The free parameters in the fit were as follows: the number of signal counts, the number of events from the flat background and the number of counts under the additional background peaks (scenarios 2 and 3).
The energy resolution was fixed to $\sigma$=1.8$\,$keV.
It was evaluated on the 511$\,$keV peak and on the two $^{60}$Co peaks at 1173$\,$keV and 1332$\,$keV that are visible in the CUORICINO energy spectrum (see Figure~\ref{fig:cuoricino_spectrum}), and it was found to be comparable for all three peaks.
A summary of the best-fit values for the $2\nu\beta\beta$ searches is reported in Table~\ref{tab:2nufit}.
Systematic uncertainties were evaluated by repeating the fitting procedure with different background models, fitting ranges and energy resolutions, and, compared to statistical uncertainties, they were found to be negligible.

\begin{table}[tb]
\begin{center}
\begin{tabular}{ccc}
&\\
\hline\hline
Scenario & $N_{S}$ & $N_{B}$\\
& [counts] & [counts/keV]\\
\hline
1 &  1.1$\pm$1.4$\pm$0.29 & 0.12$\pm$0.03\\
2 & -0.4$\pm$6.6$\pm$2.6  & 6.31$\pm$0.41\\
3 & -3.0$\pm$6.8$\pm$2.8  & 6.73$\pm$0.33\\
\hline\hline
\end{tabular}
\end{center}
\caption{2$\nu$ analysis best-fit values for the number of signal ($N_{S}$) and background ($N_{B}$) counts.
For $N_{S}$, both statistical and systematic uncertainties are reported.}
\label{tab:2nufit}
\end{table}

\section{Results}
For each decay mode and for each of the three scenarios, the posterior probability density function (p.d.f.) for the number of signal counts, $P(N_{S})$, was extracted using a Bayesian approach and assuming flat priors in the physical region ($N_{s} > 0$).
For the 0$\nu$ decay mode, because there was no evidence of a signal and the background was negligible, a Poisson p.d.f. for zero observed events was assumed for all three scenarios.
For the 2$\nu$ decay mode, the p.d.f.s were obtained as a result of the maximum likelihood fits on the spectra shown in Figure~\ref{fig:spectra}.
For each decay mode, a global p.d.f. for the decay rate was obtained as the product of the three individual p.d.f.s, $P_{TOT}(\Gamma) = \prod_{i}{P_{i}(\Gamma)}$.
In this formula $P_{i}(\Gamma) = P_{i}(N_{S})\cdot \varepsilon_{i}\cdot N(^{130}Te)\cdot t$, where the index $i$ runs over the three scenarios and $\varepsilon_{i}$ is the corresponding detection efficiency from Table~\ref{tab:cuts}.
Systematic uncertainties were included in the $P_{i}(\Gamma)$ according to the procedure described in~\cite{Andreotti:2010vj}.
This resulted in the following half life lower limits:

$$
\begin{array}{c}
T_{1/2}(2\nu\beta\beta^{*}) > 1.3 \cdot 10^{23}\,\mathrm{y}, \; 90\%\, \mathrm{C.L.}\\
T_{1/2}(0\nu\beta\beta^{*}) > 9.4 \cdot 10^{23}\,\mathrm{y}, \; 90\%\, \mathrm{C.L.}
\end{array}
$$

These new limits represent an improvement of almost 2 orders of magnitude, for both the 0$\nu$ and 2$\nu$ processes, with respect to the results of past experiments.
It is worth noting that the new lower limit on the half life of the 2$\nu$ decay mode is close to the upper bound of the theoretical calculation presented in Table~\ref{tab:prevpub}.
A more clear picture will be available once CUORE, CUORICINO's successor~\cite{Arnaboldi2004}, comes online.
This is due to the increase in target mass and improved background reduction that will be achieved in CUORE.

\section*{Acknowledgments}
The CUORICINO Collaboration owes many thanks to the Directors and Staff of the Laboratori Nazionali del Gran Sasso over the years of the development, construction and operation of CUORICINO, and to the technical staffs of our Laboratories.
In particular we would like to thank R. Gaigher, R. Mazza, P. Nuvolone, M.~Perego, B.~Romualdi, L.~Tatananni and A.~Rotilio for continuous and constructive help in various stages of this experiment.
We are grateful to our colleagues Y.G.~Kolomenski and L.~Zanotti for help and fruitful discussions.
The CUORICINO experiment was supported by the Istituto Nazionale di Fisica Nucleare (INFN), the Commission of the European Community under Contract No. HPRN-CT-2002-00322, by the U.S. Department of Energy under Contract No. DE-AC03-76-SF00098, and DOE W-7405-Eng-48, and by the National Science Foundation Grant Nos. PHY-0139294 and PHY-0500337.

\bibliography{\jobname}

\end{document}